\documentclass[aps,amssymb,12pt,showpacs,showkeys,letterpaper]{revtex4}
\usepackage{graphicx}
\usepackage{amsmath}
\usepackage{epsfig}
\usepackage{bm}

\newcommand{\be}{\begin{equation}}
\newcommand{\ee}{\end{equation}}
\newcommand{\beq}{\begin{eqnarray}}
\newcommand{\eeq}{\end{eqnarray}}

\def\J{\tilde{J}}
\def\L{\tilde{L}}

\def\r{\tilde{r}}

\usepackage{amsmath,amssymb}

\begin{document}

\title{Effective Potential Structure of the BTZ Black Hole in Rainbow Gravity }

\author{ Carlos Leiva }
 \email{cleivas@uta.cl}
\affiliation{\it Departamento de F\'{\i}sica, Universidad de
Tarapac\'{a}, Casilla 7-D, Arica, Chile}

\author{Ignacio Espinoza}
 \email{iaespino@gmail.com }
 \affiliation{\it Departamento de F\'{\i}sica, Universidad de
Tarapac\'{a}, Casilla 7-D, Arica, Chile}

\date{\today}

\begin{abstract}

  In this paper we study  the effective potential structure of the BTZ black hole in rainbow gravity for a massive and a massless particles.
\end{abstract}



\maketitle
\section{\label{sec:Int}Introduction}

Today there are many efforts  in Theoretical Physics that try to
combine quantum theory and general relativity. Several lines of
research have been developed but  none of them is completely
successful in obtaining a complete description of the quantum
gravity realm. Meanwhile, some phenomenological approaches have
been put on the table. One of them is the modification of the
dispersion relation $E^2-p^2=m^2$ \cite{piran}, with a non linear
version instead. It is very probable, after the most of the
results, that the linear version of the relation linking energy
and momenta is just a first approximation to a real non lineal
one.

On the other hand, some data seem to invite to introduce a minimal
length in physical theories. Indeed, there already exist  well
theoretically established theories, such  as String Theory or Loop
Quantum Gravity, that have some fundamental quantities: the Planck
longitude $l_p=\sqrt{\hbar G/c^3}$, the associated time scale
$t_p=l_p/c$ and the Planck energy $E_p=\hbar /t_p$. All of them
suppose that beyond these thresholds, the physics should change
dramatically.

Some proposals  to modify the Lorentz boosts through are the
approximations called Double Special relativity (DSR)
\cite{mag,giovanni2,kow, Bruno}. These theories are based on a
generalization of Lorentz transformations through a more broad
point of view of a kind of conformal transformations
\cite{Leiva1},  where there are  two observer independent scales,
velocity of light and Planck length. These theories are rather
polemical, but they are of increasing interest  because they can
be useful as effective new tools in gravity theories for example,
in Cosmology as an alternative to inflation \cite{mof,alb}, or in
other fields like propagation of light \cite{ku}, that are
related, for instance, to cosmic microwave background radiation.

There are several works about the   so called rainbow gravity,  as
an example \cite{Lsv} and references there in, but the history
begins more or less with a treatment done in
Ref.\cite{Magueijo:2002xx}.

In this paper we review the structure of effective potential of a
BTZ black hole, motivated by the fact that black holes provide
gravity conditions  to test quantum effects due to the discrete
nature of spacetime or the existence of a limit in the energy that
a particle can bear.

The paper is organized as follows: In the next section we recall
the effective potential treatment in a BTZ black hole. In the
third section we introduce the rainbow hypothesis , and in the
fourth one we discuss conclusions.

\section{\label{sec:Dilatonic}BTZ effective potential}

The metric of a BTZ black hole \cite{BTZ, cruz} is:

\begin{equation}
ds^2=-N^2dt^2+N^{-2}dr^2+r^2\{N^\phi dt+d\phi\}^2,
\end{equation}

where $N^2$ and $N^\phi$ are defined as:

\begin{equation}
N^2=-M+\frac{r^2}{l^2}+\frac{J^2}{4r^2};   \qquad
  N^\phi=-\frac{J}{2r^2}
\end{equation}

Here, $-\infty < t < \infty$,  $-\infty < r < \infty$ and $0 < t <
2\pi$.

M is the mass and J the angular momentum of the black hole. In
order to apply the effective potential method,we define de
Lagrangian of a particle in this metrics as.

\begin{equation}
\mathcal{L}=
-N^2\dot{t}^2+N^{-2}\dot{r}^2+r^2\{N^\phi\dot{t}+\dot{\phi}\}^2,\label{lag}
\end{equation}

where dots indicate derivation with respect to a affine parameter
$\lambda$.

We can find two movement integrals in the Lagrangian {\ref{lag}}.
Making the variations in $t$ and in $\phi$ respectively we get:

\begin{equation}
\mathcal{E}=\{-M+\frac{r^2}{l^2}\}\dot{t}+\frac{J}{2}\dot{\phi},
\end{equation}

\begin{equation}
\mathcal{J}=r^2\dot{\phi}-\frac{J}{2}\dot{t}.
\end{equation}

Is worth to mention that $\mathcal{E}$ can not be identified
directly with the energy of the particle because the metric is not
flat when $r$ tends to infinite. But, we will call it energy
because it comes from the variation of the Lagrangian respect to
the coordinate time.

Furthermore, it is possible find another movement constant,
$ds^2=\mathcal{L}=-\xi^2$. Where $\xi^2$ has the values $1$ for
timelike geodesics (massive particles) and $0$ for null geodesics
(massless particles).

Using the results of \cite{cruz}, we can calculate the velocity in
terms of two potentials:

\begin{equation}
\dot{r}^2=(\mathcal{\tilde{E}}-V^+_{eff})(\mathcal{\tilde{E}}-V^-_{eff}),
\end{equation}

where $\mathcal{\tilde{E}}$ is defined as
$\mathcal{\tilde{E}}=\mathcal{E}/\sqrt{M}$, and the effective
potentials are:

\begin{equation}
V_{eff}=\frac{\J \L}{2\r^2}\pm
\frac{1}{\r^2}\sqrt{4\xi^2\r^4(-1+\r^2)+\J^2(\L^2+\xi^2\r^6)},
\end{equation}

where

\begin{equation}
\L=\frac{L}{l\sqrt{M}}, \qquad \J=\frac{J}{lM}, \qquad
\r=\frac{r}{lM},
\end{equation}


\section{\label{sec:Dilatonic}BTZ effective potential in Rainbow Gravity }

Let's introduce now the functions $f(\tilde{\mathcal{E}})$ and
$g(\tilde{\mathcal{E}})$ in the metric. These functions are analog
to  those introduced by \cite{mag,giovanni2,kow, Bruno} in the
nonlinear version of the dispersion relation
$E^2/f^2=p^2/g^2+m^2$, but we are going to choose they in a way
that doesn't modify the speed of light , that is to say $f=g$.
Moreover, we claim that these functions tend to $1$ when
$\tilde{\mathcal{E}}\ll \tilde{\mathcal{E}}_p$ where
$\tilde{\mathcal{E}}_p$ can be identified with a fundamental
length scale $l_p=1/\tilde{\mathcal{E}}_p$. So we choose $f(E)$
as:

\begin{equation}
f(E)=\frac{1}{1-\tilde{\mathcal{E}}/\tilde{\mathcal{E}}_p}
\end{equation}

Then, the Lagrangian is modified in the following way.

\begin{equation}
\mathcal{\hat{L}}=
\frac{-N^2}{f(\mathcal{E})^2}\dot{t}^2+\frac{N^{-2}}{f(\mathcal{E})^2}\dot{r}^2+r^2\{\frac{N^\phi}{f(\mathcal{E})}\dot{t}+\frac{\dot{\phi}}{f(\mathcal{E})}\}^2,\label{lag}
\end{equation}

Now, we can find  two movement integrals for this new Lagrangian:

\begin{equation}
\hat{\mathcal{E}}=\frac{1}{f^2}\{-M+\frac{r^2}{l^2}\}\dot{t}+\frac{J}{2f^2}\dot{\phi}
\end{equation}

and

\begin{equation}
\hat{\mathcal{L}}=\frac{r^2}{f^2}\dot{\phi}-\frac{J}{2f^2}\dot{t}
\end{equation}

Now, we can find an equation for the velocity in terms of
effective potentials, in an analog way to the normal case, with
the re scaled variables,

\begin{equation}
\hat{\Omega}=\frac{\hat{\mathcal{E}}}{\sqrt{M}}, \qquad
\hat{\Lambda}=\frac{\hat{\mathcal{L}}}{l\sqrt{M}},
\end{equation}

and using re scaled variables of the last section the velocity is
expressed as:

\begin{equation}
\dot{r}^2=(\hat{\Omega}-V^+_{eff})(\hat{\Omega}-V^-_{eff}),\label{eqb}
\end{equation}

where

\begin{equation}
\hat{V}_{eff}=\frac{\hat{\Lambda}\J}{2\r^2}\pm
\frac{1}{2f\r^2}\sqrt{f^2\hat{\Lambda}^2(\J^2-4\r^2+4\r^4)+4\xi^2(\J^2-\r^4+\r^6)}
\end{equation}

The plot of the original and new effective potentials for a
massive particle and for a massless particle  are depicted in
Fig.1:

\begin{figure}[!h]
  \begin{center}
  \includegraphics[width=170mm]{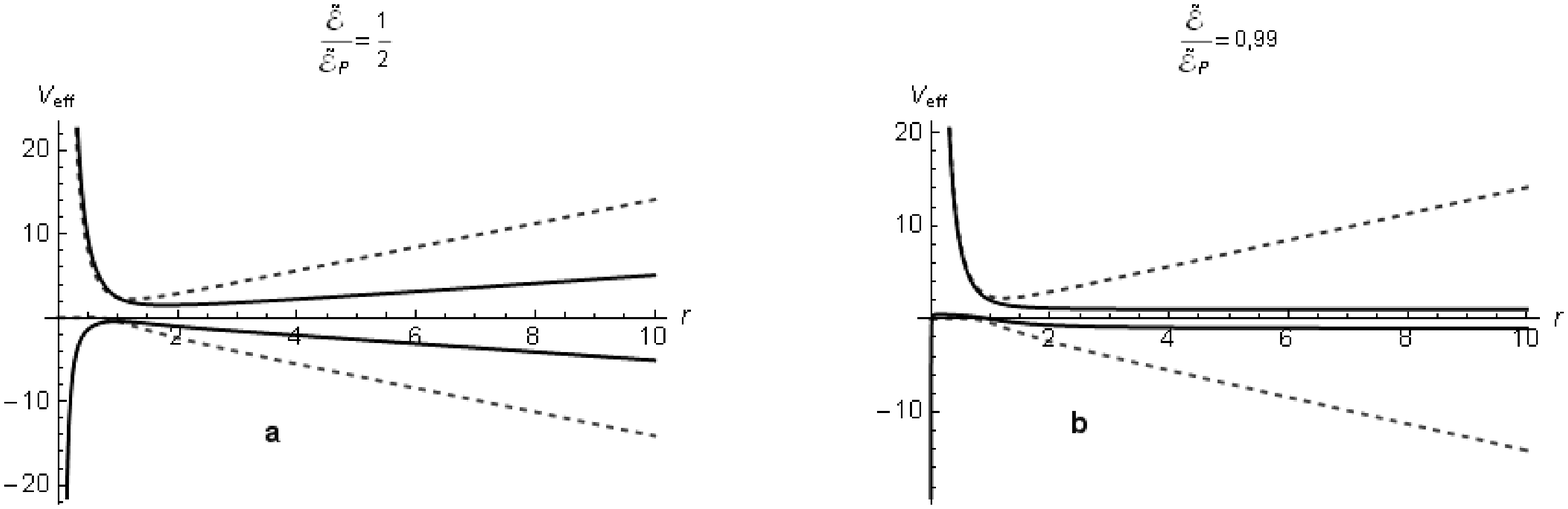}
  \end{center}
  \caption{This plot shows the effective potential for a massive particle in  the non modified metrics (dotted line) and
  in the modified one (continuous line) , when $\tilde{\mathcal{E}}/\tilde{\mathcal{E}}_p=1/2$ (a) and when $\tilde{\mathcal{E}}/\tilde{\mathcal{E}}_p=0,99$ (b).}
  \label{fig:trng}
\end{figure}

In Fig. 1 we can see that the potentials become flat when
$\tilde{\mathcal{E}}/\tilde{\mathcal{E}}_p\rightarrow 1$, so a
massive particle can now become free for a suitable distance far
from the black hole. Furthermore, the zone where the velocity is
non real in eq \ref{eqb}, is reduced almost to zero, even it's not
totally null.

The plot of the original and new effective potentials for a
massless particle and for a massless particle  are depicted in
Fig.2

\begin{figure}[!h]
  \begin{center}
    \includegraphics[width=170mm]{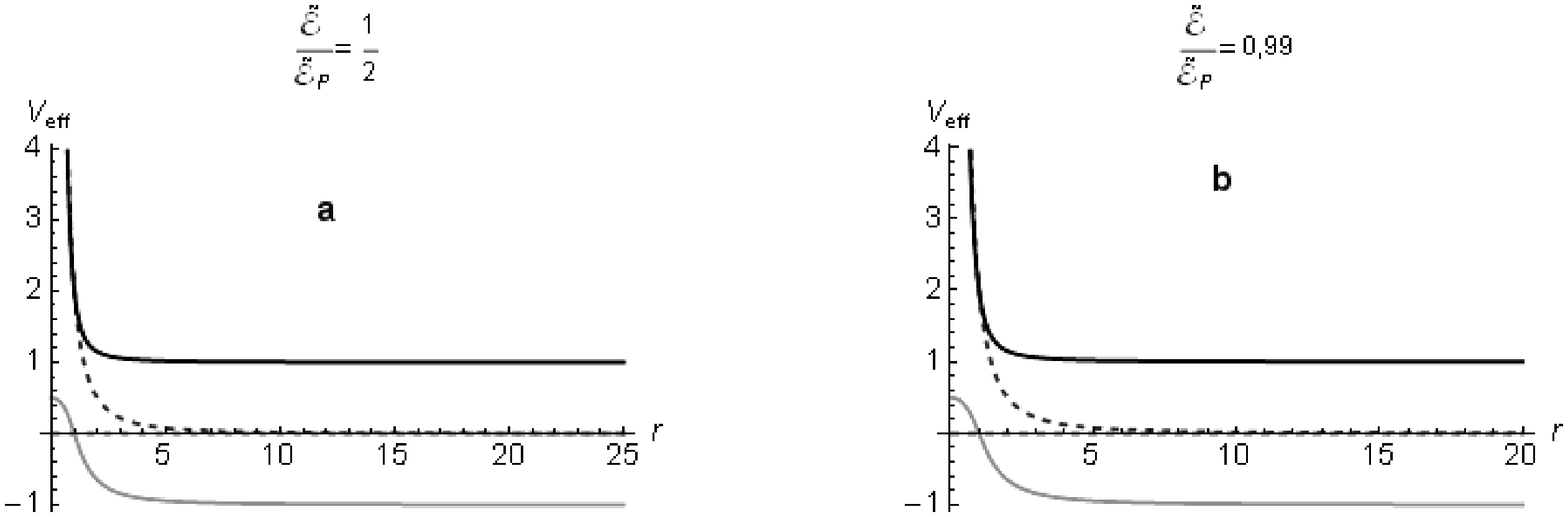}
  \end{center}
  \caption{This plot shows the effective potential for a massive particle in  the non modified metrics (dotted line) and
  in the modified one (continuous line) , when $\tilde{\mathcal{E}}/\tilde{\mathcal{E}}_p=1/2$ (a) and when $\tilde{\mathcal{E}}/\tilde{\mathcal{E}}_p=0,99$ (b).}
  \label{fig:trng}
\end{figure}

In Fig. 2 we can see that the potentials is flat when
$\tilde{\mathcal{E}}/\tilde{\mathcal{E}}_p\rightarrow 1$, as they
are in the non modified case, but a massive particle has a not
allowed energies where the velocity is non real in eq \ref{eqb},
broader than in the original case.

\section{\label{sec:Dilatonic2}Discussion and Outlook}

In this work we have seen that the introduction of functions $f$
has a very impressive consequences. The structure of the orbits
allowed, due the deformation of the effective potentials  for a
massless particle and the ligated state for a massive particle are
very distorted from the classic case. These results lead to think
that it is worth to investigate about another kinds of black
holes, specially in 3-D searching for observable perturbations in
cosmological issues.

\begin{acknowledgments}
 C.L. was
supported by Grant UTA  N$^0$ 4722-09.
\end{acknowledgments}

\end{document}